\documentclass{sbrt2018eng}

\ifdefined\usePortuguese
\usepackage[brazil]{babel}   
\usepackage[latin1]{inputenc}
\else
\fi

\ifdefined\appendices
\else
\ifdefined\addAppendixAsPreamble 
\usepackage[titletoc,title]{appendix} 
\else
\usepackage{appendix} 
\fi
\fi

\usepackage{makeidx}  
\usepackage{graphicx}
\usepackage{color} 
\definecolor{darkgreen}{rgb}{0,0.35,0}

\ifdefined\keywords
\else

	\def\keywords{\vspace{.5em}{\bfseries\textit{Index Terms}---\,\relax%
	}}
	\def\endkeywords{\par}
\fi

\usepackage{url} 

\ifdefined\RCSfile 
\else
\usepackage{cite} 
\fi

\usepackage{multirow} 

\usepackage[bbgreekl]{mathbbol}

\ifdefined\amsmath
\else
\usepackage{amsmath}
\fi

\ifdefined\amsfonts
\else
\usepackage{amsfonts}
\fi

\usepackage{amssymb}

\ifdefined\proof
\else
\usepackage{amsthm}
\fi

\ifdefined
\else
\ifdefined\akbook
\ifdefined\lulu
\typearea[6mm]{1} 
\else
\usepackage[a4paper, top=3cm, bottom=3.4cm, left=1.8cm, right=2.5cm]{geometry}
\fi
\fi
\fi

\usepackage{mathrsfs} 
\usepackage{makeidx}  
\usepackage{graphicx,color}

\usepackage{multirow} 
\usepackage{listings} 
\usepackage{ifpdf} 

\ifpdf

\ifdefined\hyperref
\else
\usepackage[pdftex,colorlinks]{hyperref}
\fi

\usepackage{epstopdf}
\hypersetup{%
pdftitle={Some title},
pdfauthor={Aldebaro - LASSE - UFPA},
pdfkeywords={DSP,Signal},
pdfstartview={FitH}, 
urlcolor=black,
linkcolor=black,
citecolor=black,
}

\fi 

\begin{document}

\title{Evaluation of Simplified Methodology for Obtaining mmWave MIMO Channels from Ray-Tracing Simulations}

\author{Isabela Trindade, Brenda Vilas Boas and Aldebaro Klautau 
\thanks{Isabela Trindade, Brenda Vilas Boas and Aldebaro Klautau are with 
LASSE - Telecommunications, Automation and Electronics Research and Development Center, Bel\'em-PA, Brazil, E-mails: isabela.trindade@itec.ufpa.br, and \{brendavb,aldebaro\}@ufpa.br. 
}}

\maketitle

\markboth{XXXVI SIMPÓSIO BRASILEIRO DE TELECOMUNICAÇÕES E PROCESSAMENTO DE SINAIS - SBrT2018, 16-19 DE SETEMBRO DE 2018, CAMPINA GRANDE, PB} {XXXVI SIMPÓSIO BRASILEIRO DE TELECOMUNICAÇÕES E PROCESSAMENTO DE SINAIS - SBrT2018, 16-19 DE SETEMBRO DE 2018, CAMPINA GRANDE, PB}

\begin{abstract}
The use of higher frequencies and MIMO is important in many 5G use cases.  However, the available channel models for millimeter waves (mmWaves) 
currently demand investigation and the number of measurements is still limited.
Using simulators is a current practice in mmWave MIMO research and ray tracing is considered one of the most accurate techniques. Due to the relatively long time of ray tracing simulations, it is common practice to adopt a simplified simulation methodology in which omnidirectional antennas are simulated and, later, the results are used together with a geometrical model to consider that antenna arrays were used. This allows flexibility and decreases the overall time spent with simulations. This paper investigates the corresponding assumptions and how accurate are the results of the simplified methodology when compared to effectively using antenna arrays in the ray tracing simulation. The preliminary results indicate that the distance between transmitter and receiver needs to be sufficiently large.

\end{abstract}


\section{Introduction}

Many applications of mmWaves are on the way: 5G, wireless local area networks, vehicular area networks, wearables, etc. Despite the large bandwidth availability, the use of higher frequencies imposes many challenges which makes the study of mmWaves channels relevant 
\cite{heath_overview_2016}. 
Given the small wavelength, compact antenna arrays are feasible and most mmWave communication systems will use MIMO techniques to enhance capacity at both the transmitter and receiver. Therefore, a simplified realistic representation of the mmWave channel is critical to design a robust mmWave application \cite{heath_overview_2016}.

Virtual channel model \cite{sayeed_deconstructing_2002}, also called angular model \cite{tse_fundamentals_2005} or geometric channel model \cite{klautau_5g_2018}, is an intermediate channel representation between statistical models and physical models that has proven to ease the understanding of scattering on channel capacity \cite{sayeed_deconstructing_2002,heath_overview_2016}. 
The work in \cite{heath_overview_2016} gives a brief introduction on virtual representation of MIMO channels for studying signal processing techniques on mmWaves. In \cite{klautau_5g_2018} and \cite{va_inverse_2017} the narrowband virtual channel version is used to generate a database for vehicle to infrastructure (V2I) environment which was applied to a machine learning method for beam selection on the first article and used on the second for beam alignment through multipath fingerprint.  

This article presents a systematic study of MIMO channels using ray-tracing (RT) software which evaluates 
the simplified and comprehensive MIMO channel generation proposed in \cite{sayeed_deconstructing_2002}. 
When a MIMO simulation is done at a RT software, like Remcom InSite, the output channel matrix is dependent of antenna array size and orientation; 
 therefore, the study is very dependent of the scenario created and requires a new simulation every time one desires to change antenna orientation or array size. The simplified methodology 
enables simulations on omnidirectional mmWaves scenarios that are stored and used to generate channel matrices in a post-processing phase, allowing flexibility when deciding antenna array size and orientation given that the visualization of clusters 
provides better insight to define antenna orientation. 

This article is structured as follows, section \ref{sec:methods} describes the methods used to generate the channel matrices. Section \ref{sec:results} presents the simulation results for the simplified model and a complete MIMO simulation at InSite RT software
; besides, the two methods are compared. Finally, section \ref{sec:conclusions} presents the conclusions of the ongoing study.







\section{Methods}
\label{sec:methods}


RT 
can provide very accurate results for 5G scenarios ~\cite{Rappaport14,stabler_mimo_2009_ak}. However, its computational cost increases exponentially with the maximum
allowed number of reflections and diffractions~\cite{arikawa_simplified_2014}. Besides, RT simulations are \emph{site specific}, depending on the specific propagation environment. In this regard, it is important to develop a general simulation approach for RT to be able to reuse simulation datasets, reducing investigation time. Therefore, 
this section will present the methodology for simplified MIMO simulation and all-MIMO simulation using a RT simulator. In both cases, we consider a single cell with single user MIMO. A urban canyon V2I scenario was build as in \cite{klautau_5g_2018} at 60 GHz frequency. 








\subsection{Full RT MIMO Simulation}
\label{sec:full}
For full RT MIMO simulation, all the available features of the RT software are used to generate a MIMO channel matrix. In our case, we set InSite as in the general scenario description above, and choose the antennas at transmitter and receiver as uniform linear arrays (ULA) 
with $N_\mathrm{tx}$ and $N_\mathrm{rx}$ elements, respectively. With this configuration, a new simulation need to be run every time a different array size is chosen.



\subsection{Simplified RT Simulation for MIMO}
\label{sec:simp}
To increase flexibility, all receivers and  transmitters use half-wave omnidirectional antennas in this RT simulation approach. The role played by MIMO antenna arrays is added in a post-processing stage based on the geometric channel model which  
allows, for example, to simulate arrays with various number of antennas without the need to re-run 
the relatively long RT simulations. 
The information collected from the RT software 
for each transmitter / receiver pair $(m,n)$ are: average time of arrival $\overline \tau_{mn}$, 
total transmitted $\hat P_\textrm{tx}$ and received $\hat P_\textrm{rx}$ powers, and
for the $\ell$-th ray, $\ell=1,\ldots,L$, complex channel gain $\alpha_{\ell}$, time of arrival $\tau_\ell$, angles $\phi_\ell^D$, $\theta_\ell^D$, $\phi_\ell^A$, $\theta_\ell^A$, corresponding respectively to azimuth and elevation for departure and arrival. 
In this modeling strategy, the $L$ most prominent rays for each transmitter / receiver pair $(m,n)$ informed by the RT simulator is taken in account to compute the MIMO channel $\mathbf{H}_{mn}$ for each pair $(m,n)$.


For the post-processing method, the \emph{narrowband geometric channel model} is applied on the RT collected data and the MIMO channel $\hat{\mathbf{H}}_{mn}$ is calculated as


\begin{align}
\hat{\mathbf{H}}_{mn} = \sqrt{N_\mathrm{tx} N_\mathrm{rx}}\sum_{\ell = 1}^L \alpha_{\ell} \mathbf{a}_r(\phi_\ell^A, \theta_\ell^A)\mathbf{a}^H_t(\phi_\ell^D, \theta_\ell^D), 
\end{align}
where $N_\mathrm{tx}$ and $N_\mathrm{rx}$ are the numbers of array antenna elements at transmitter and receiver, $\alpha_\ell$ is the complex channel gain, $\mathbf{a}_r(\phi_\ell^A, \theta_\ell^A)$ and $\mathbf{a}_t(\phi_\ell^D, \theta_\ell^D) $ are the steering (column) vectors at the receiver and transmitter for the $\ell$-th path, respectively, and $(\cdot)^H$ denotes the Hermitian. The resulting MIMO channel matrix $\hat{\mathbf{H}}_{mn}$ has dimension $N_\mathrm{rx} \times N_\mathrm{tx}$ and a rank that depends on the angular spacing among the rays.
In this paper we are not considering the variation of $\hat{\mathbf{H}}_{mn}$ over time.


\section{Results}
\label{sec:results}

\begin{figure}
\centering
\includegraphics[width=0.8\linewidth]{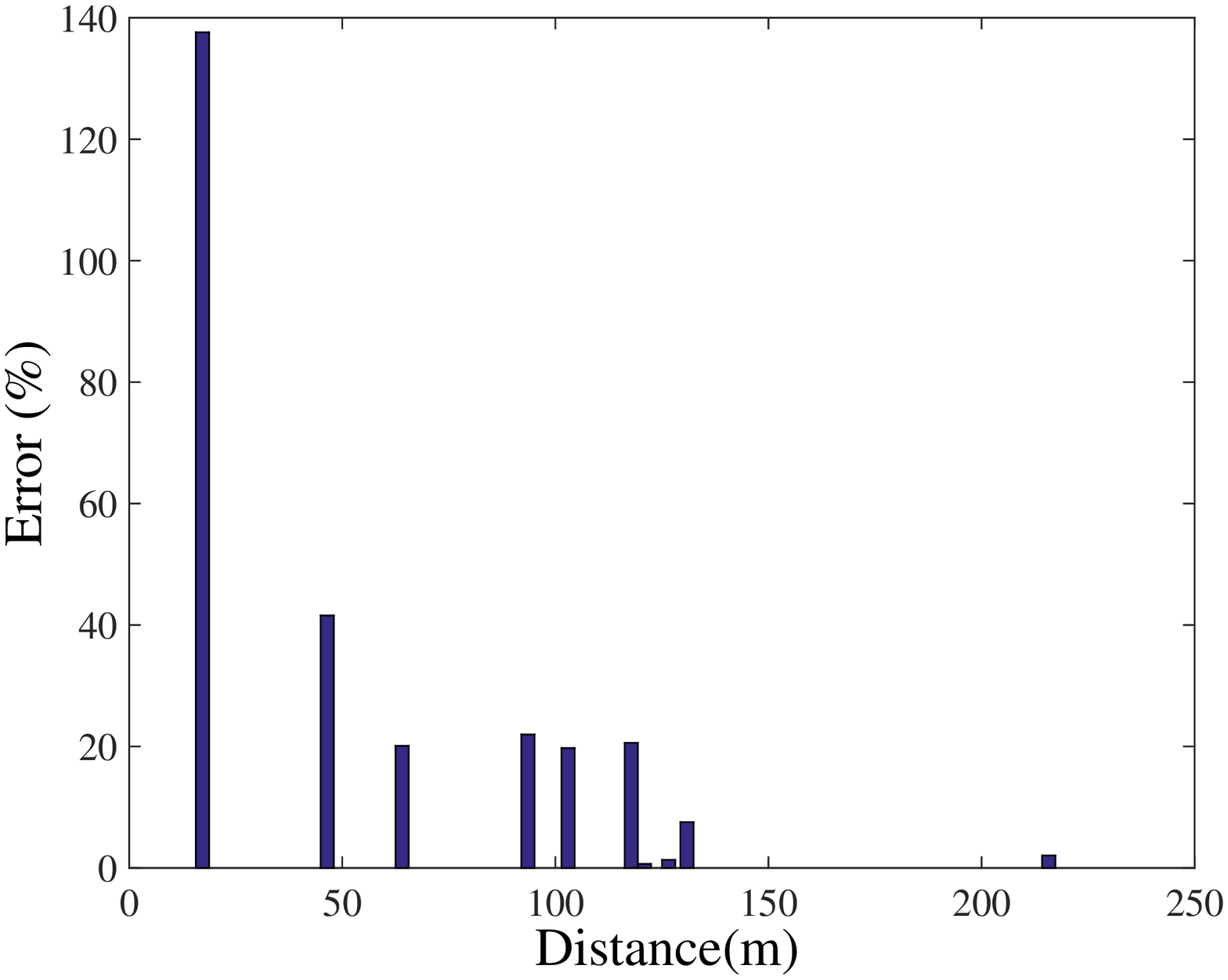}
\caption{Bar distribution of the Errors}
\label{fig:rxbarsdist}
\end{figure}

The full MIMO simulation at RT used an ULA of 4 antenna elements in each transmitter and receiver, other parameters were set as in Section \ref{sec:full}. The simplified channel methodology was conducted as described in Section \ref{sec:simp}.  
Fig. \ref{fig:rxbarsdist} presents the error between the two channel calculation methods. Analyzing the difference between the channel matrices studied,
 it was observed that the closest receiver from the transmitter had the largest error 
whereas the receiver that obtained the least error was at NLOS condition. 
Then, it could be inferred that it is less difficult to estimate the channel from the receiver far away from the transmitter than one which is closer.

\begin{figure}
\centering
\includegraphics[width=0.8\linewidth]{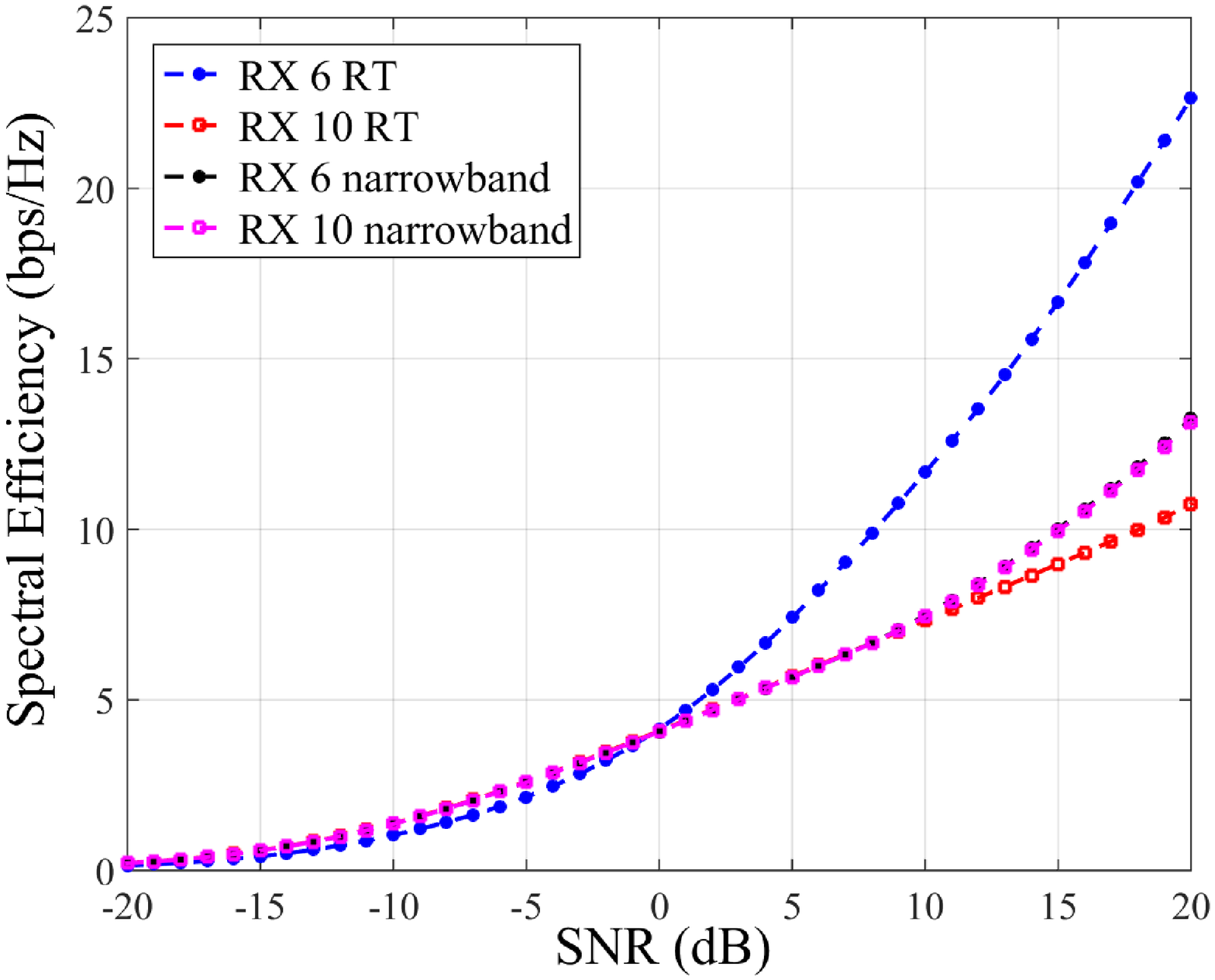}
\caption{Capacity of the MIMO channels produced by RT software and narrowband model.}
\label{fig:capacity}
\end{figure}

Fig. \ref{fig:capacity} presents the capacity for the two cases explained here, the closest LOS receiver (RX 6) and the NLOS receiver (RX 10) which presented the lowest error percentage. It can be inferred that the channel approximation is still valid for low signal do noise ratio (SNR) regime. 

\section{Conclusions}
\label{sec:conclusions}
This paper discusses a modeling problem of interest to modern mmWave MIMO systems. The preliminary results indicate that for long distances between the transmitter and receiver the simplified model gives similar results as the full simulation, but care must be exercised when using the simplified methodology in ray tracing simulations. The continuation of this work includes a systematic assessment of the accuracy of the methodology with respect to the estimation error of the corresponding channel matrices and on figures of merit such as spectral efficiency.


\bibliographystyle{IEEEtran}

\bibliography{references}

\end{document}